\documentstyle[prl,aps,floats]{revtex}

\def\[{\begin{equation}}
\def\]{\end{equation}}

\def\gtil{\widetilde {g}}
\def\ktil{\widetilde {k}}
\def\Qtil{\widetilde{Q}}
\def\qtil{\widetilde {q}}
\def\Gtil{\widetilde  { G}}

\def\0til{\bar {0}}

\def\tanh{{\rm tanh}}
\def\cosh{{\rm cosh}}
\def\ln{{\rm ln}}

\input epsf.tex
\begin{document}
\author{C. Bourbonnais and B. Dumoulin} \address{Centre de Recherche en Physique du Solide et 
D\'epartement de Physique,}
\address{Universit\'e de Sherbrooke, Sherbrooke, Qu\'ebec, Canada J1K 2R1.}
\title{{\bf Theory of Lattice and Electronic Fluctuations
 in  Weakly Localized Spin-Peierls Systems}}
\date{July 1996}
\maketitle

\begin{abstract}
 A theoretical approach to the influence of one-dimensional lattice fluctuations
on electronic properties in weakly localized spin-Peierls systems is proposed
using the renormalization group and the functional integral techniques.
 The interplay between the renormalization group flow of correlated   electrons
and one-dimensional  lattice fluctuations is taken into account by 
 the one-dimensional functional integral method  in the adiabatic
limit. Calculations of spin-Peierls precursor effects  on  response functions are
carried out explicitely and the prediction  for the temperature dependent 
magnetic susceptibility  and  nuclear relaxation is compared with  available
experimental data for (TMTTF)$_2$PF$_6$. 
  
\end{abstract}

\section{INTRODUCTION }
Among the possible structural phase transitions in low-dimensional
systems,  the
spin-Peierls (SP) instability has recently received particular attention.  
\cite{schegolev,Bray}.
 In contrast with the Peierls transition, whose normal phase is metallic,
 pronounced insulating properties and antiferromagnetic spin correlations are
found   far ahead of the SP ordering temperature. However, owing to
the quasi-one-dimensional character  of these systems, precursor
effects 
should  appear in  both types of transitions. Since spin correlations are
essentially one-dimensional in character,
  lattice fluctuations should emerge as diffuse scattering in x-ray experiments,
and as a  so-called pseudo-gap effect in  magnetic
properties. The pseudo-gap effect is well known to be the
hallmark of strongly anisotropic  Peierls systems
\cite{schegolev,Lee}, but     surprisingly,  its observation  is a rather
uncommon  feature of most SP systems   and the underlying explanation  for
this is not really known so far\cite{Bray}. Correspondingly, the incitement
to achieve a  theoretical description of fluctuation effects in the SP 
case   did not receive any experimental justification, so that  most 
theoretical efforts were rather  devoted to the description of the ordered
state\cite{Bray,Cross-Fisher,fukuyama}.   However,  recent observations of  
SP  fluctuation  effects in  the organic  systems
(TMTTF)$_2$X and (BCPTTF)$_2$X with X=PF$_6$, AsF$_6$, has prompted a renewed
interest  for this  problem
\cite{pouget,creuzet,Liu,CaronCB,prl}. In both cases, one-dimensional SP
precursors are clearly seen in structural and magnetic measurements well above
the true SP ordering temperature.  

This paper will be entirely devoted to the theoretical description
of one-dimensional lattice and electronic correlations in SP  systems of
the (TMTTF)$_2$X series. Unlike (BCPTTF)$_2$X, which are already good insulators
at room temperature, (TMTTF)$_2$X are weakly localized  and still show a metallic
behaviour in this temperature range. It is only  around 
$T_\rho
\simeq  220$K that carriers  undergo a Mott-Hubbard localization, below which 
 the resistivity becomes thermally activated\cite{laversanne}.  
  Albeit the magnetic susceptibility remains unaffected at $T_\rho$, NMR
spin-lattice relaxation rate measurements have revealed the existence of strong
1D antiferromagnetic spin correlations below this temperature
range\cite{creuzet,wzietek2}.  Below 60K, one-dimensional SP lattice softening
is  borned out by x-rays and  has a marked influence on the temperature
dependence of both susceptibility and nuclear relaxation\cite{pouget,creuzet}.
In order to tackle the problem of one-dimensional  electronic and lattice
fluctuations effects in SP systems we shall combine the renormalization
group technique for fermions and  the functional-integral
method\cite{CaronCB,prl}. A similar approach has been recently applied -with some
success- to strongly localized SP systems of the (BCPTTF)$_2$X series \cite{prl}.
Since the itineracy of carriers is  much more pronounced in the high
temperature domain of (TMTTF)$_2$X series,   both spin {\it and} charge
degrees of freedom must be considered in the renormalization flow, which in
practice makes the extension of the previous approach 
slightly more complicated.

\section{The model and the Partition function}  

In  setting up the essential ingredients of the model, 
 it is first convenient  to study the  partition function of  the
electron-acoustic-phonon interaction model, namely when  direct electron-electron
interaction  is absent. The corresponding hamiltonian is well known  to take the
form
\cite{SSH-Barisic}:
\begin{eqnarray}
H=  \sum_{p,k,\sigma} \epsilon_p(k) c^\dagger_{p,k,\sigma}c_{p,k,\sigma} +
      \sum_{Q} \omega(Q) (b^\dagger_Qb_Q + {1\over 2}) +
(L)^{-1/2}\sum_{\{p,k,Q,\sigma\}}g(k,Q)c^\dagger_{p,k+Q,\sigma}c_{-p,k,\sigma}
(b^\dagger_{-Q}+ b_Q),
\end{eqnarray}
where $\epsilon_p(k)=v_{F}(pk-k_{F})$ is  the linearized spectrum around the 1D
Fermi points  $p k_{F}=\pm k_{F}$  for   free fermions  and $\omega
(Q) = \omega_{D} | \sin (Qa/2) |$ is the dispersion relation for the 
phonons 
 ($\hbar=1$). Here $v_F$ is the Fermi velocity  and $\omega_D$
is the Debye frequency, which coincides with the phonon
energy at $Q=\pm 2k_F (=\pm \pi/a$,  when the band is  half-filled). As for the
interacting part, the electron-phonon coupling constant is given by:
\begin{eqnarray}
 g(k,Q) = i4{g_{a}\over{\sqrt{2M\omega (Q)}}}\sin
(Qa/2)\cos (ka + Qa/2), 
\label{ephconst}
\end{eqnarray}  where $M$ is  the molecular mass and $g_{a} = -dt_{a}/dx >0 $
is the spatial modulation of the longitudinal hopping integral $t_a$. 
The dynamics of phonons  being purely harmonic, their
integration in the partition function $Z= {\rm Tr}
e^{-\beta H}$
   can be carried out exactly, yielding an
effective retarded interaction among electrons $(\beta=1/T, k_B=1)$. The result
can be put in the following functional-integral form   
\begin{equation}
Z=\ \int [d\psi^*][d\psi] \ e^{S_{\rm e}^0[\psi^*,\psi]}\ \exp
\bigl(-\int_0^\beta d\tau_1
 d\tau_2\  {\cal H}(\tau_1-\tau_2)\bigr),
\end{equation}
over the anticommuting Grassman variables $\psi^{(*)}$. The free electronic
action  is given by 
\begin{eqnarray}
S_{\rm e}^0[\psi^{*},\psi ]= \int_0^\beta
d\tau\ 
\sum_{p,k,\sigma}\psi^{*}_{p,\sigma}(k,\tau)[G^0_p(k,\tau)]^{-1}
\psi_{p,\sigma}(k,\tau),
\end{eqnarray}
 where $G^0_p(k,\tau)= [-\partial/\partial\tau
-\epsilon_p(k)]^{-1}$ is the free electron propagator. As for the retarded
interaction ${\cal H}(\tau_1-\tau_2) $, we will proceed  to
 a ``g-ology'' decomposition which  only retains $2k_F$ phonon exchange
between electrons at $\pm k_F$, with the result 
\begin{eqnarray}
{\cal H}(\tau_1-\tau_2)=  && \
(2L)^{-1}\sum_{\lbrace p,k,q,\sigma\rbrace}\ g_{\rm 1,ph}(\tau_1-\tau_2)\
\psi^{*}_{p,\sigma_{1}}(k_{1}+ p2k_F +q,\tau_1)
\psi^{*}_{-p,\sigma_{2}}(k_{2}-p2k_F-q,\tau_2)\cr 
&& \hskip 5 truecm \times \ 
\psi_{p,\sigma_{2}}(k_{2},\tau_2) \psi_{-p,\sigma_{1}}(k_ {1},\tau_1)\cr
 +&&  \ (2L)^{-1}\sum_{\lbrace p,k,q,\sigma \rbrace } \ g_{\rm
3,ph}(\tau_1-\tau_2)\ 
\psi^{*}_{-p,\sigma_1}(k_{1}+p2k_F-q- pG,\tau_1)
\psi^{*}_{-p,\sigma_2}(k_{2}
-p2k_F+q,
\tau_2)\cr
&& \hskip 5 truecm \times\  \psi
_{p,\sigma_2}(k_{2},\tau_2)\psi_{p,\sigma_1}(k_{1},\tau_1), 
\end{eqnarray}
where $G=4k_F$ is a reciprocal lattice vector. The retarded backscattering and
umklapp coupling constants 
 are given by
\begin{eqnarray}
g_{\rm 1,ph}(\tau_1-\tau_2) = && g(-pk_F,p2k_F)g(pk_F,-p2k_F)
D^0(2k_F,\tau_1-\tau_2)\cr
= && {8 g_a^2\over M\omega_D}D^0(2k_F,\tau_1-\tau_2),\cr 
g_{\rm 3,ph}(\tau_1-\tau_2) = &&-\eta g_{\rm 1,ph}(\tau_1-\tau_2),
\nonumber 
\end{eqnarray}
where $D^0(2k_F,\tau_1-\tau_2)= -[e^{-\omega_D \mid\tau_1-\tau_2\mid}
+2(e^{\beta\omega_D}-1)^{-1} \cosh\bigl(\omega_D(\tau_1-\tau_2)\bigr)]$ is the
phonon propagator at
$2k_F$. 
Here, $\eta \le 1$ is a positive constant that reflects the half-filled character
of the band. For a single half-filled band,
$\eta=1$ while for   systems like  (TMTTF)$_2$X  with a 
quarter-filled band and a small dimerization gap, one can take $\eta\ll
1$\cite{BB}.  From above, one observes  that, for exchange of  $2k_F$ acoustic
phonons,
$g_{\rm 3,ph}$ and $g_{\rm 1,ph}$ are opposite in sign. The retarded interaction
can then be written uniquely in terms of composite fields:
\begin{eqnarray}
{\cal H}(\tau_1-\tau_2)= \int dx\ \sum_{M} g^M_{\rm ph}(\tau_1-\tau_2)
O^{M*}(x,\tau_1)O^M(x, \tau_2),
\end{eqnarray}
 where  we have introduced the composite field 
$$
O^{M*}(x,\tau)= {1\over 2}\bigl(O^*(x,\tau) + MO(x,\tau)\bigr),
$$
with
$$
O(x,\tau)=   \sum_\sigma\psi^*_{-,\sigma}(x,\tau)\psi_{+,\sigma}(x,\tau).
$$
 These corresponds to  ``site'' ($M=+$) and ``bond"  ($M=-$)
charge-density-wave (CDW) correlations at half-filling.  The related  
combinations of couplings  are given by
$$
g^M_{\rm ph}(\tau_1-\tau_2)= g_{\rm 1,ph}(\tau_1-\tau_2) + Mg_{\rm
3,ph}(\tau_1-\tau_2).
$$ 
In order to analyze these correlations, we apply  an
Hubbard-Stratonovich transformation on the retarded part of the
interaction 
\begin{eqnarray}
Z=  \int  [d\phi^*][d\phi]\  [d\phi^{M}]\    e^{S_e^{0}[\psi*,
\psi] } 
 \ && \exp
\Bigl\{-\int dxd\tau_1d\tau_2
           \sum_{M=\pm} \phi^{M}(x,\tau_1)\mid  g^M_{\rm
ph}(\tau_1-\tau_2)\mid^{-1}
 \phi^{M}(x,\tau_2) \Bigr\} \cr
 && \times  \exp \Bigl\{-\int dx d\tau\sum_{M=\pm}
\lambda_M O^{M}(x,\tau)\phi^{M}(x,\tau) \Bigr\}, 
\end{eqnarray}
where  $\phi^{\pm}$ are  real auxiliary fields
for site 
and bond  fluctuations respectively, and $\lambda_{+}=2,\lambda_{-}=
2i$.  It is worth noting that, in the absence of umklapp scattering,
 $g^{-}_{\rm ph}=g^{+}_{\rm ph}$  and there is no
difference between bond and site density wave fluctuations
 so that  both auxiliary fields 
will combine to produce amplitude and phase fluctuations of the complex field
$\phi= \phi^+ + i\phi^- = \mid \phi \mid e^{i\varphi}$, as found in 
incommensurate Peierls systems.

 Unretarded repulsive electron-electron interaction at half-filling
will promote bond with respect to site electronic correlations.  
These essential ingredients of the SP  instability  can be added to the 
model by following  the ``g-ology"  prescription \cite{Solyom}, in which  the
direct  interaction among right- and left-moving carriers is decomposed 
in     terms of backward ($g_1$), forward ($g_2$) and ($g_3$) umklapp scattering
coupling constants. The full  partition function of the model in the
Fourier-Matsubara space  then becomes 
\begin{eqnarray}
Z[h^M]= &&\int   [d\psi^*][d\psi][ d\phi^M] \  \exp\Big\lbrace S_{e}^0[\psi^*,
\psi] + S^0[\phi^M]+    S_\lambda[\psi^*,
\psi,\phi^M]+ S_I[\psi^*,\psi]+ S_h[\psi^*,\psi,h^M]\Big\rbrace\cr
 \equiv && \int   [d\psi^*][d\psi][ d\phi^M]  \   \exp\Big\lbrace
\sum_{p,\ktil,\sigma}[G^0_p(\ktil)]^{-1}\psi^{*}_{p,\sigma}(\ktil)
\psi_{p,\sigma}(\ktil) -\sum_{\Qtil}\mid\phi^M(\Qtil)\mid^2
\mid g^M(\omega_m)\mid^{-1}
\cr &&\hskip 4truecm +\sqrt{T\over L}\sum_{\Qtil}\lambda_M
O^M(\Qtil)\phi^M(\Qtil)\cr &&  \hskip 4truecm + {T\over 2L}
\sum_{\lbrace p,\tilde{k},\tilde{Q},\sigma \rbrace }
g_{1}\psi^{*}_{p,\sigma_{1}}(\tilde{k}_{1}+\tilde{Q})
\psi^{*}_{-p,\sigma_{2}}(\tilde{k}_{2}-\tilde{Q})
\psi_{-p,\sigma_{1}}(\tilde{k}_
{1}) \psi_{p,\sigma_2}(\tilde{k}_{2})\cr   &&  \hskip 4truecm +{T\over
2L}\sum_{\lbrace p,\tilde{k},\tilde{q},\sigma
\rbrace }g_{2}
\psi^{*}_{p,\sigma_{1}}(\tilde{k}_{1}+\tilde{q})
\psi^{*}_{-p,\sigma_{2}}(\tilde{k}_{2}-\tilde{q})
\psi_{p,\sigma_{1}}(\tilde{k}_
{1}) \psi_{-p,\sigma_{2}}(\tilde{k}_{2})\cr
  &&\hskip 4truecm + {2T\over L}
\sum_{\lbrace p,\tilde{k},\tilde{Q},\sigma
\rbrace }g_{3}
\psi^{*}_{p,\sigma_1}(\tilde{k}_{1}+\tilde{Q}_p)
\psi^{*}_{p,\sigma_2}(\tilde{k}_{2}
-\tilde{Q}_p + p\tilde{G})\psi
_{-p,\sigma_1}(\tilde{k}_{1})\psi_{-p,\sigma_2}(\tilde{k}_{2})\cr
 &&\hskip 4truecm + \ \sum_{\Qtil,\mu}\Bigl( O^M_\mu(\Qtil)h_\mu^{M*}(\Qtil) +
{\rm H.c}\Bigr)
  \Big\rbrace,
\label{Z}
\end{eqnarray} 
where $\ktil=
(k,\omega_n=(2n+1)\pi T))$,
$\Qtil_p=(pQ,\omega_m=2\pi mT)$,  
$\Gtil=(4k_F,0)$,  $G^0_p(\ktil)= [i\omega_n- \epsilon_p(k)]^{-1}$, and   
$g^M(\omega_m)= 8g_a^2(M\omega_D)^{-1}(1 -M\eta) D^0(2k_F,\omega_m)$. The
integration  measures are $[d\psi^*][d\psi]= \prod_{p\ktil,\sigma}
d\psi^*_{p,\sigma}(\ktil) d\psi_{p,\sigma}(\ktil)$ and $[
d\phi^M]=\prod_{M,\Qtil >0}
\mid~g^M_{\rm ph}(\omega_m)~\mid^{-1}d\phi^M(\Qtil)d\phi^M(-\Qtil)$ for fermion
and auxiliary fields, respectively. Here 
\begin{eqnarray}
D^0(2k_F,\omega_m)= {-2\omega_D \over \omega^2_m + \omega^2_D},
\end{eqnarray}
corresponds to the bare phonon propagator.  Finally, $S_h$ 
is an additional term   which couples   an infinitesimal source field $h_\mu^M $
to site  bond  ($M=\pm$,
$\mu=0$) CDW  and site ($M=+$, $\mu=1,2,3$) SDW correlations. These correspond
to composite fields defined by $O^{M\ast}_{\mu}(\Qtil)={1 \over 2}
\bigl(O^{\ast}_{\mu}(\Qtil) +MO_{\mu}(\Qtil)\bigr)$, with 
$O^{\ast}_{\mu=0,1,2,3}(\Qtil)=(T/L)^{1/2}\sum_{\ktil,\alpha,\beta}
   \psi^{\ast}_{+,\alpha}(\ktil + \Qtil) \sigma^{\alpha\beta}_{\mu}
   \psi_{-,\beta}(\ktil)$, $\sigma_{1,2,3}$ and $\sigma_0$ being the Pauli and
the identity matrices respectively. In the following the source-field term will
be useful  for the calculation of relevant  response functions for the
SP  instability.

\section{Renormalization group results}

 The problem of low-energy electronic
and lattice correlations  reduces   to the study of 
interacting electrons  coupled to the  fluctuating field
$\phi^M$, which can be done by first applying the  renormalization group 
approach  developped in Ref.\cite{CBLG,prl} for the fermion degrees of
freedom. It  consists in first integrating high-energy  fermion states, namely 
we write
$\psi^{(*)}\to \psi^{(*)} +
\bar{\psi}^{(*)}$, where  $\bar{\psi}^{(*)}$ describes degrees of
freedom to be integrated over in the outer energy shell of tickness
${1\over 2}E_0(\ell)d\ell$ on both sides of the Fermi level at $\pm k_F$  and for
all
$\omega_n$. Here $E_0(\ell)= E_0e^{-\ell} $ is the band energy cut-off at the
step
${\ell}$ and $E_0\equiv 2E_F$ is the initial band width, which is twice the
Fermi energy.   Keeping the
$\phi^M$'s fixed, this is formally written  as 
\begin{eqnarray}
 Z[h^M] \propto && \int_<   [d\psi^*][d\psi][ d\phi^M]\ 
e^{S[\psi^*,\psi,\phi^M,h^M]_\ell - \beta{\cal F[\phi^M]}_\ell}
\int_{o.s} [d\bar{\psi}^*][d\bar{\psi}]\  
e^{\bar{S}^0_e[\bar{\psi}^*\bar{\psi}]}\ \Bigl(e^{
\bar{S}_\lambda
+\bar{S}_I +
\bar{S}_h}\Bigr)\cr        =
&& \int_<   [d\psi^*][d\psi][ d\phi^M]\ 
e^{S[\psi^*,\psi,\phi^M,h^M]_\ell - \beta{\cal F[\phi^M]}_\ell}
\exp{\Bigl(\sum_n {1\over n!}
\langle(\bar{S}_\lambda + \bar{S}_I+ \bar{S}_h)^n\rangle_{o.s}\Bigr)} \cr
\propto &&
\int_<   [d\psi^*][d\psi][ d\phi^M]\ 
e^{S[\psi^*,\psi,\phi^M,h^M]_{\ell+d\ell}-
\beta{\cal F}[\phi^M]_{\ell + d\ell}},
\label{trace}
\end{eqnarray}
which leads to a recursion relation for electronic parameters of the
action including those related to source fields for the calculation of
response function (see below) and for 
${\cal F}[\phi^M]_\ell$, which is a free-energy functional which collects
all the  contributions in  the auxilliary field $\phi^M$. At finite
temperature, the partial integration is conducted down to $\ell_T=
\ln(E_F/T)$. In the present renormalization group procedure, electronic degrees
of freedom are treated in the continuum limit so that the  discreteness of the
underlying lattice will be neglected  in the following
\cite{prl}.    

\subsection{Electronic part}
 
\subsubsection{Couplings and one-particle self-energy} 
 It is useful for the following discussion to  recall the well-known
two-loop RG results for the purely electronic part ($\phi^M=0$) in zero field
($h^M=0$). The effect  of the lattice
fluctuations  and the calculation of relevant response functions  will be
considered  afterwards.  In this purely electronic limit, 
 the single-particle fermion propagator transforms  according to the recursion
relation 
$z_1(\ell+d\ell)[G_p^0]^{-1}=z_1(\ell)z_1(d\ell)[G_p^0]^{-1}$. At two-loop level
the outer-shell correction to $z_1(d\ell)$ comes  from the $n=2$ $\langle
\bar{S}_{\rm I}^2\rangle_{o.s}$ terms of (\ref{trace}), which yields  the flow
equation:
\begin{equation}
 {d \ln z_1^{-1}\over d\ell} =
-\frac1{16}\bigl\{[2\gtil_2(\ell)-\gtil_1(\ell)]^2 -\gtil_3^2(\ell) +
3[\gtil_1^2+3 \gtil_3^2]\bigr\},
\label{z1}
\end{equation}
where the $\gtil_i(\ell)\equiv g_i(\ell)(\pi v_F)^{-1}$ are the
normalized coupling constants at
$\ell$. The latter transform as $\gtil_i(\ell+ d\ell)=\gtil_i(\ell)
z_1^{-2}(d\ell)z_{2,3,4}(d\ell)$, which, together with $z_1$, are obtained from
one-loop
$\langle \bar{S}_{\rm I}^2\rangle_{o.s}$ and two-loop $
\langle \bar{S}_{\rm I}^3\rangle_{o.s}$ outer shell corrections
to two-particle four-point vertex functions $\Gamma_{1,2,3}(\ell+d\ell)=
z_{2,3,4}(d\ell)\Gamma_{1,2,3}(\ell)$.  
 This is  known to yield the flow equations \cite{kimura,Solyom,CBLG} 
\begin{eqnarray}
{d\gtil_1 \over d\ell} &&= -\gtil^{2}_1 -{1 \over 2}\gtil^{3}_1, \cr
 {d(2\gtil_2 -\gtil_1) \over d\ell} && = \gtil^{2}_3 
[1-{1 \over 2}(2\gtil_2 -\gtil_1)], \cr
 {d\gtil_3 \over d\ell} &&= \gtil_3 (2\gtil_2 -\gtil_1)
[1-{1 \over 4}(2\gtil_2 -\gtil_1)]-{1 \over 4}\gtil^3_3, 
\label{couplings}
\end{eqnarray}
 For (TMTTF)$_2$X compounds,  $\gtil_1\simeq \gtil_2 > 0$ 
 are repulsive at $\ell=0$, whereas a  finite
dimerization of the TMTTF stacks can be parametrized by a small and
positive $\gtil_3 \ll \gtil_1$ \cite{BB,CaronCB}.  One then gets the
inequality 
 $\gtil_1-2\gtil_2 < \gtil_3 $  indicating that  umklapp scattering is
relevant. The flow of charge couplings $2\gtil_2-\gtil_1$ and $\gtil_3$
then scales towards strong coupling  while the spin coupling $\gtil_1$
is found to be  marginally irrelevant 
 with the fixed point values $\gtil_3^*(\ell\rightarrow \infty) \rightarrow 2$,
$\gtil_2^*(\ell\rightarrow \infty)
\rightarrow 1$, and $\gtil_1^*\rightarrow 0$. 
The strong coupling sector of the RG flow is reached when
$g_3(\ell\equiv\ell_\rho)
\approx 1
$, where $\ell_\rho=\ln(E_F/T_\rho)$ defines  the temperature scale for the
presence of  a charge  (Mott-Hubbard)  gap
$\Delta_\rho\equiv 2T\rho$. Neglecting transients  between weak and
strong coupling regimes, one-particle self-energy corrections resulting from the
solution of Eq. (\ref{z1})
 can be put in the following  scaling form
\begin{equation}
z^{-1}_1(\ell) \approx z_1^{-1}(\ell_\rho)
\bigl(E_0(\ell)/\Delta_\rho)^{\theta_\rho^*},
\label{scaling1}
\end{equation}
which  depicts the reduction of the quasi-particle weight at the Fermi level
with 
$\theta^*(g_1^*,g_2^*,g_3^*)= 3/4$ being evaluated   at the fixed
point.
 
\subsubsection{Response functions}
 The pertinent  electronic response functions involved in the description of the
SP  instability in one dimension are those related to site SDW
($\mu\ne0,M=+$) and bond CDW ($\mu=0,M=-$) correlations\cite{CaronCB}. They can
be  readily calculated via the renormalization of the source field term  
$S_h$ \cite{CBLG} which,  at scale $\ell$, reads 
\begin{eqnarray}
S_h[\psi^{\ast},\psi, h^M]_{\ell} =
\sum_{\{\mu,M,\Qtil\}}
\bigl\{ z^{M}_{\mu}(\ell)
h^{M\ast}_{\mu}(\Qtil)O^{M}_{\mu}(\Qtil) + {\rm H.c}. 
 -\chi^{M}_{\mu}(2k_F,\ell)h^{M\ast}_{\mu}(\Qtil) 
h^{M}_{\mu}(\Qtil)\bigr\},
\end{eqnarray}
where $z^{M}_{\mu}(\ell)$ is the renormalization factor for the pair vertex
part, while 
\begin{equation}
\chi^{M}_{\mu}(2k_F ,\ell)= 
-(\pi v_F)^{-1}\int^{\ell }_0 \bar {\chi}^{M}_{\mu}(\ell')d\ell',
\label{response}
\end{equation}
is the  $2k_F$ response function and 
$\bar{\chi}_\mu^M= (z^{M}_{\mu})^2$ is the auxiliary susceptibility in the
$(\mu,M)$ channel considered . The one- and two-loop corrections to
$z^{M}_{\mu}$ come 
from the outer shell averages $\langle\bar{S}_h\bar{S}_I\rangle_{\rm o.s}$ and
$\langle \bar{S}_h\bar{S}_I^2\rangle_{\rm o.s}$ for $n=2$ and $n=3$
respectively. This leads to the flow equation 
\begin{equation}
{d\ln\bar {\chi}^{M}_{\mu} \over d\ell}=\gtil^{M}_{\mu}(\ell) - {1 \over
2}\bigl\{\gtil^{2}_1(\ell)+
\gtil^{2}_2(\ell)-\gtil_1(\ell)\gtil_2(\ell) 
+{1\over 2}\gtil^{2}_3(\ell) \bigr\},
\label{auxilliaire}
\end{equation}
where $\gtil^{\pm}(\ell)=\gtil_2(\ell) \mp \gtil_3(\ell)-2\gtil_1(\ell)$ for
the CDW channel and
$\gtil^{+}_{\mu \not=0}(\ell)=\gtil_2(\ell)+ \gtil_3(\ell)$ for the SDW channel.
Following the example of (\ref{scaling1}) for the one-particle self-energy
renormalization,   
$\bar{\chi}_\mu^M$  can be expressed in the following scaling form
\begin{eqnarray}
\bar{\chi}^{M}_{\mu}(\ell)\approx &&
  \bar{\chi}^{M}_{\mu}(\ell_\rho) \bigl(E_0(\ell)/
\Delta_\rho\bigr)^{-\gamma^{\ast M}_{\mu}},
\label{scaling2}
\end{eqnarray}
where $\bar{\chi}^{M}_{\mu}\bigl(\ell_\rho)$ is the weak coupling contribution
 below the charge gap. The exponents
$\gamma^{\ast+}_{1,2,3}=\gamma^{\ast-}= 3/2$ obtained by evaluating the
r.h.s of (\ref{auxilliaire}) at the fixed point indicate that only
site SDW and bond CDW correlations are singular at half-filling.  It should be
mentionned, however,  the 
$g_i^*$ obtained at two-loop level are well known to overestimate the values of
the exponents. Higher order contributions are expected to bring them closer to
the exact value
\begin{equation}
\gamma^{\ast+}_{1,2,3}=\gamma^{\ast-}\equiv \gamma^*=1,
\end{equation}
which is known to result  from  more elaborate
calculations at $\ell \gg \ell_\rho$ \cite{VoitRev}. 

Before closing this section, it is useful for applications (see \S III.B.5 and
\S IV.B) to compute the imaginary part of the retarded response function at low
(real) frequency. From  Ref. \cite{CBLG},  it is 
directly connected to 
$\chi_\mu^M$ as follows: 
\begin{equation}
{\rm
Im}\chi_\mu^M(2k_F +q,\omega) = \bar{\chi}_\mu^M(T)\ {\rm
Im}\chi^0(q+2k_F, \omega),\qquad \omega \to 0
\label{imaginaire}
\end{equation} 
for $v_Fq\ll T$,  where  
\begin{eqnarray}
{\rm Im}\chi^0(q+2k_F, \omega )= &&{2T\over L}{\rm
Im}\sum_{k,\omega_n} G^0_-(k,\omega_n)G^0_+(k+2k_F +q,\omega_n +\omega +i0^+ )\cr
  =\ &&-{ 8 \omega\over{Tv_F\cosh^{2}(\beta
v_{F}q/4)}}, \qquad \omega\to 0,
\label{nue}
\end{eqnarray}
is the imaginary part of the low-frequency non-interacting response function
near
$2k_F$ (see Appendix A.2).

\subsection{Influence of lattice fluctuations}
 \subsubsection{Ginzburg-Landau functional}

 When the lattice auxiliary field
$\phi^M$ is  taken into account, the partial integration  (\ref{Z})
generates a  series of terms in $n\ge 2$ powers of the
$\phi^M$'s via  the closed fermion loops $\langle
\bar{S}_{\lambda}^n\rangle/n!$. This will not only lead to corrections for
$S^0[\phi^M]_\ell$ for
$n=2$, but it yields  a recursion relation for the  quantum
Ginzburg-Landau  free-energy functional ${\cal F}[\phi^M]_\ell$,   when combined
to the
$n>2$ terms
\cite{prl,CBLG}.  Since  only
$2k_F$ bond correlations are singular in the CDW channel for repulsive
couplings, only the  dependence on the bond field $\phi^-$ needs to be retained,
with the result  
\begin{eqnarray}
\beta{\cal F}[\phi^-]_{\ell +d\ell}= \beta{\cal F}[\phi^-]_{\ell}\  -\ 
\sum_{n\ge 2}\sum_{\Qtil} B_n(d\ell)\  
\phi^-(\Qtil_1)\ldots\phi^-(\Qtil_n)\delta_{\Sigma_{\tilde{Q}}=\Gtil}.
\end{eqnarray}
The  quadratic ($n=2$)  and quartic ($n=4$) outer shell contributions lead  to 
\begin{eqnarray}
B_2(d\ell)= &&
(\pi v_F)^{-1}[z^-(\ell)]^2\ d\ell, \cr \cr
 B_4(d\ell)\simeq && - {T\over L}{7\zeta(3)[z^-(\ell)]^4\over 2\pi^3
v_F [E_0(\ell)]^2}\ d\ell,
\label{param}
\end{eqnarray} 
where  $\zeta(3)\simeq 1.2\ldots$.
For the problem at hand, one can assume  
that adiabaticity between electrons and the lattice is sufficiently strong that
quantum lattice corrections  can be neglected, thereby  allowing 
the static limit. Up to the quartic term, one obtains  the following Ginzburg-Landau free energy functional 
(in
real space)
\begin{eqnarray}
{\cal F}[\phi^-]_{\ell_T}=  \int dx \ \bigl[\  a(T)     \bigl(\phi^-(x)\bigr)^2 
+ c
\left({d\phi^-\over dx}\right)^2 + b \bigl(\phi^-(x)\bigr)^4\ \bigr],
\label{GL}
\end{eqnarray}  
at temperature $T$, where $a(T)$, $c$ and $b$ are the Ginzburg-Landau
parameters.  The coefficients $a(T)$ and $c$ of the quadratic term are
respectively   given by 
\begin{eqnarray}
a(T)= &&[g_{\rm ph}^-(0)]^{-1} +\chi^-(
2k_F,T)\cr
\approx && a'(T/T_{SP}^0-1)\cr
c  \ \approx &&{1\over 2} a'(v_F/\pi T^0_{SP})^2,
\end{eqnarray}
where $a'=(\pi v_F)^{-1} \bar{\chi}^-(T_{SP}^0)$.
 Here the bond CDW response function $\chi^-(2k_F,T)$ in Eq. (\ref{response}),
together with (\ref{scaling2}), have been used in the linearization of 
$a(T)$  around  the SP  mean-field temperature 
\begin{equation}
T_{SP}^0 \approx T_\rho\bigl(\mid \gtil_{\rm ph}^-\mid
\bar{\chi}^-(T_\rho)\bigr)^{1/\gamma^*}.
\label{mftemp}
\end{equation}
for $T_\rho \gg T_{SP}^0$. It is worth noting that
since
$\mid g_{ph}^-\mid \  \propto
g_a^2$, then 
$T_{SP}^0\propto \ g_a^2 $, which turns out to be the same power dependence
on the  electron-phonon coupling constant  that was found in previous
mean-field calculations on more localized systems \cite{Cross-Fisher}. As for
the rigidity parameter  $c$, it is readily obtained  in Fourier space from the
$Q=2k_F + q$ expansion of
$\chi^{-}(Q,T_{SP}^0)$   using  the approximate expression 
$$ 
\ell (T_{SP}^0, 2k_F + q)= \ln\left\{ {E_F \over T_{SP}^0}[ 1+
(v_Fq/\pi T_{SP}^0)^2]^{-{1\over2}}
\right\},
$$
as boundary condition of the logarithmic integration in 
(\ref{response}) near $T_{SP}^0$. This form essentially  coincides
  with the $Q$ dependence of the elementary Peierls bubble near $2k_F$. 
Finally, the flow of the mode-mode coupling in (\ref{param}) is
stopped at
$\ell_{T_{SP}^0}$ and reads
\begin{eqnarray}
b\  &&\approx  (\pi v_F)^{-1}{7 \zeta (3) \over 16(1+\gamma^*) (\pi 
T^{0}_{SP})^2}  [\bar
{\chi}^-(T^0_{SP})]^2, 
\end{eqnarray}
for $T_\rho \gg T_{SP}^0$. 

Fluctuation effects in $\phi^-$ below $T^0_{SP}$  are then governed by  the
classical functional integral ${\cal Z}= \int[d\phi^-]
\exp(-\beta {\cal F}[\phi^-])$,  which can be carried  out exactly 
using the transfer matrix method \cite{diet-sca}. An important quantity
to compute is  the static 1D SP  response function $\chi_{SP}$.
From the results of Ref.\cite{diet-sca}, one finds
\begin{eqnarray}
\chi_{SP}(2k_F +q,T)= && {1\over T}\int dx \  \langle\langle
\phi^-(x)\phi^-(0)\rangle\rangle\  e^{-i(q+2k_F)x}\cr
 = && {2 \beta \ \langle\langle \ (\phi^-)^2 \ \rangle\rangle \  \xi_{SP}
\over 1 + q^2 \xi^{2}_{SP}},
\end{eqnarray}
where 
$$
\langle\langle \ (\ldots)\  \rangle\rangle = {\cal Z}^{-1}\int[d\phi^-]\ 
(\ldots)
\exp(-\beta {\cal F}[\phi^-]), 
$$
denotes a   statistical average over $\phi^-$. Here $\xi_{SP}$ is the 
correlation length of  the  real order field $	\phi^-$, which grows
exponentially below
$T^{0}_{SP}$ according to 
\begin{eqnarray}
\xi_{SP}= 
 { T\over \rho}\sqrt{c\over\mid a\mid} \  e^{\rho/ T} ,
\end{eqnarray}
where
\begin{eqnarray}
\rho = {(2c)^{1 \over 2 }\mid a \mid^{3 \over 2} \over b}
\approx  3.8T^{0}_{SP},
\end{eqnarray}
for $T\ll T_{SP}^0$.  The temperature profile of both $\xi_{SP}$ and the mean
square fluctuations  $\langle\langle (\phi^-)^2 \rangle\rangle$ are plotted in
Figure 1.

{\epsfxsize 7cm
\begin{figure}
\centerline{\epsfbox{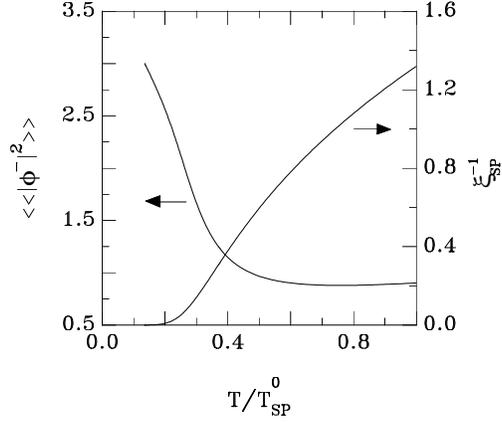}} 
\caption{Temperature variation of the mean square of the SP order parameter
(left scale )  and the inverse of the correlation length (right scale)
normalized by the lattice constant.}
\end{figure}}

\subsubsection{Pseudo-gap effect} 

The influence of the static
SP  lattice fluctuations governed by (\ref{GL}) on
 one-particle electronic properties at the step
$\ell_T$ of the RG procedure   will be analyzed by considering  to leading order
the  one-electron self-energy contribution of Figure 2\cite{Lee}, which  can be
obtained from (\ref{Z})  after 
 an integration over the $\phi^-$ field. The result is  
  \begin{eqnarray}
\Sigma^*_p (\ktil, \lbrace \phi^- \rbrace ) = -TL^{-1}z_1^{-1}z_0^{-}
\sum_{q}G_{-p}^{0}(k -p2k_{F}-q,i\omega_{n})\chi_{SP}(2k_{F} + q,T).
\label{self}
\end{eqnarray} 
The bare propagator appearing in $S_{\rm e}^0$  then becomes 
\begin{eqnarray}
[G_{p}(\ktil, \lbrace \phi^- \rbrace)]^{-1} = && z_1[G^0_p(\ktil)]^{-1}
-\Sigma_p^*(\ktil,\lbrace \phi^- \rbrace)\cr
= && z_1(i\omega_{n} -
\epsilon_{p}(k)) - {z_1^{-1}(z^-_0)^{2}\ \langle\langle
|\phi^-|^{2}\rangle\rangle\over i\omega_{n} +
\epsilon_{p}(k) + iv_{F}\xi_{SP}^{-1}(T)}.
\label{propag2}
\end{eqnarray}
{\epsfxsize 7cm
\begin{figure}
\centerline{\epsfbox{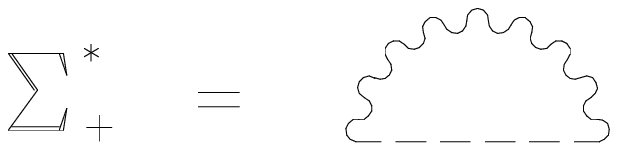}}
\caption{Leading one-particle self-energy correction due to SP fluctuations.
for electrons at
$+k_F$. The wiggly (dashed)  line corresponds to the lattice field
($-k_F$) propagator
$\chi_{SP}$.
 }
\end{figure}}
A relevant quantity to compute is the reduction of the  density of states per
spin resulting from SP  fluctuations 
(pseudo-gap effect).  After 
analytic continuation of $ [G_p]^{-1}$ to real
frequencies,  this  is found to be 
\begin{eqnarray}
 D[\omega,\lbrace \phi^- \rbrace] = && - { z_1\over\pi  L}\sum_{p,k}   {\rm
Im} G_{p}(k,\omega,\lbrace \phi^- \rbrace)\cr
 = &&
 {1\over\pi v_F}  {\alpha [2(\vartheta +
\kappa]^{1/2}\over{[2(\vartheta + \kappa) - \alpha^{2}]{\vartheta}}},
\label{dos}
\end{eqnarray}
where the presence of the renormalization factor $z_1$ ensures that the
pseudo-gap effect is the result of lattice fluctuations.  Following the notation
of Ref.\cite{Lee}, we have 
\begin{eqnarray}
&& \alpha = v_{F}\ \xi_{SP}^{-1}\ \langle\langle
|\Phi^-|^{2}\rangle\rangle^{-1/2}\cr && \tilde{\omega} =  \omega \
\langle\langle |\Phi^-|^{2}\rangle\rangle^{-1/2}\cr && {\kappa} = 1 +
{1\over{4}}\alpha^{2} -
\tilde{\omega}^{2} \cr && {\vartheta} = ({\kappa }^{2} +
\tilde{\omega}^{2}\alpha^{2} )^{1/2},
\label{definition}
\end{eqnarray}
where we have defined $|\Phi^-|^{2}=
z_1^{-2}(z^-)^2 \mid\phi^-\mid^2$. Therefore, as the temperature decreases below
the characteristic
$T_{SP}^0$, lattice fluctuations grow and will progressively freeze electronic
degrees of freedom. In turn, this reduction will affect the RG flow of various
quantities considered in \S III. The inclusion of lattice  fluctuation effects
in the RG flow will be considered  to lowest order where vertex corrections due
to exchange of  lattice fluctuations are neglected. For example, at one-loop
level of the RG,
 this amounts to substituting  $G(\ktil,\lbrace \phi^-
\rbrace)$ for one of the propagators appearing in the formal expression of the
Landau, Peierls, and Cooper  elementary susceptibilities (Figure~3 and  Appendix
A). At the two-loop level,  all the relevant next-to-leading singular 
diagrams shown in Figure~4  involve absorption and emission of an electron-hole
pair at  small
$\qtil$ in the intermediate state, which can   be dressed by lattice fluctuations
(Appendix B).
{\epsfxsize 8cm
\begin{figure}
\centerline{\epsfbox{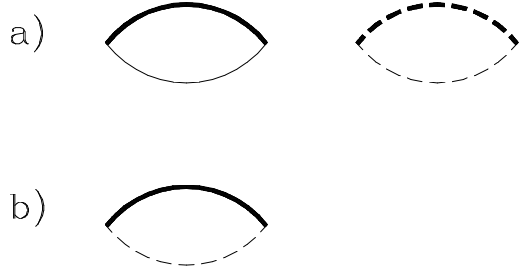}}
\caption{Dressed electron-hole bubbles in the a) Landau  and b) Peierls (or
Cooper) channels. Thick and thin continous (dashed) lines correspond to
dressed and bare electron propagators, respectively  at $+k_F$
($-k_F$). }
\end{figure} }
{\epsfysize 6cm
\begin{figure}
\centerline{\epsfbox{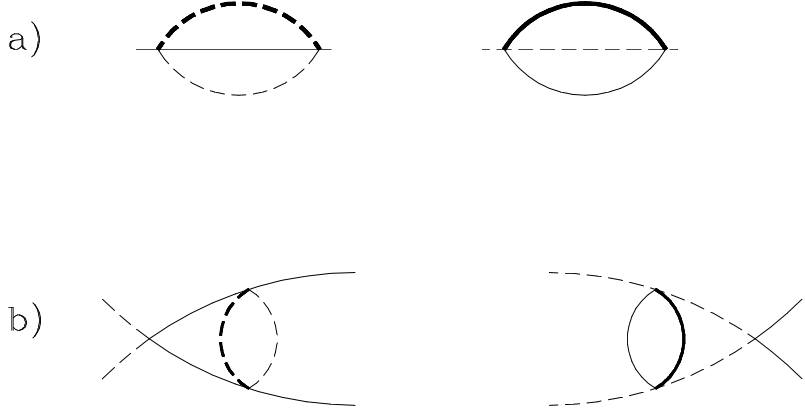}}
\caption{ Two-loop  a) one-particle self-energy  and b) four points vertex
 singular diagrams involved in  the RG  flow  including  the leading order
corrections  due to lattice fluctuations.}
\end{figure}}
From the results of
Appendix A and B, this amounts to replacing the outer shell logarithmic
contribution of all diagrams by  the following expression  
\begin{eqnarray}
d\ell
\to \pi v_F\ D[E_0(\ell)/2,\lbrace \phi^-
\rbrace] d\ell,
\label{pseudo}
\end{eqnarray}
where $D[E_0(\ell)/2,\lbrace \phi^-\rbrace]$ is density of states (\ref{dos})
 in the presence of a pseudo-gap.

 Substituting this  RG generator in
the flow equation for the  renormalization factor $z_1^{-1}$ in  (\ref{z1}) will
slow down the  electronic contribution to the decay of the quasi-particle weight
below
$T_{SP}^0$. Following the example of the purely electronic case,  the 
approximate  ``scaling" (transient-free) form (\ref{scaling1}) then becomes
\begin{equation} z^{-1}_1(T) \approx
z^{-1}_1(T_{SP}^0)
\bigl(T/T_{SP}^0)^{\theta_\rho^*(T)},
\end{equation}
where $z^{-1}_1(T_{SP}^0)$ is given by (\ref{z1}) and $\theta_\rho^*(T)=  \pi
v_F  D[T,\lbrace
\phi^-
\rbrace]\theta_\rho^*$ becomes a temperature dependent exponent below
$T_{SP}^0 $.  

\subsubsection{Staggered response functions}

The pseudo-gap will affect  electronic correlations  of the
$(\mu,M)$ channel as well. Indeed, substituting   (\ref{pseudo}) in
(\ref{auxilliaire}) yields the following approximate ``scaling" form 
\begin{eqnarray}
\bar{\chi}^{M}_{\mu}
 (T,\{\phi^-\}) \approx \bar{\chi}_\mu^{M}(T_{SP}^0) 
\bigl(T/T_{SP}^0\bigr)^{-\gamma^*(T)},
\label{chibarfl2}
\end{eqnarray}  
where $\bar{\chi}_\mu^{M}(T_{SP}^0) $ is given by (\ref{scaling2}) and 
$\gamma^*(T)=
\pi v_F  D[T,\lbrace
\phi^-
\rbrace]\gamma^*$.

As for the imaginary part (\ref{imaginaire}),   the above results and
Appendix A.2 lead to
\begin{eqnarray} {\rm Im}\chi^M_\mu(q + 2k_F, \omega,\{\phi^-\})=
&&-
 \bar{\chi}_\mu^M(T,\{\phi^-\}){\rm Im}\chi(q + 2k_F, \omega,\{\phi^-\})\cr
 \simeq &&  -\bar{\chi}_\mu^M(T,\{\phi^-\})  {\pi D[v_Fq/2,\{\phi^-\}]
\over 8T\cosh^{2}(\beta v_{F}q/4)}\ \omega,\qquad \omega\to 0
\label{imchi}
\end{eqnarray}
which is further reduced by lattice fluctuations.

\subsubsection{Magnetic susceptibility}

The calculation of the  spin response function $\chi_s$ at
 small
$(q,\omega)$  is known in the purely electronic situation 
\cite{CB2}. Its generalization when lattice
fluctuations are present is
straightforward  and yields   
\begin{equation}
\chi_s(q,\omega,\{\phi^-\})= {2 \mu_B^2\chi(q,\omega,\{\phi^-\})\over 1-
{1\over 2}g_1(T)
\chi(q,\omega,\{\phi^-\})}, 
\label{chis}
\end{equation}   
 where in the above scheme of approximation  $\chi(q,\omega,\{\phi^-\})$ is
the  elementary electron-hole bubble dressed by lattice fluctuations
(Fig. 3a and Appendix A.1), and
$g_1(T)$ is given by (\ref{couplings}) at $\ell_T$.  According to the results of
Appendix A, one finds  
\begin{eqnarray}
{\rm Re}\chi(q,\omega, \{\phi^-\})= && \ \int_{0}^{T}
D[\omega',\{\phi^-\}] \left(-{\delta n\over \delta \omega'}\right) d\omega'\sum_p
{pv_Fq\over pv_Fq-\omega},\cr
{\rm Im}\chi(q,\omega, \{\phi^-\})= && \ \int_{0}^{T}
D[\omega',\{\phi^-\}] \left(-{\delta n\over \delta \omega'}\right)
d\omega'\sum_p  p\pi v_Fq\delta(pv_Fq-\omega),
\label{chi0}
\end{eqnarray}   
 for the real and imaginary
parts of the dressed electron-hole bubble. It follows that in the static
($\omega
\to 0$) and uniform ($q\to 0$) limit, the temperature variation of the spin
susceptibility
$\chi_s(T,\{\phi^-\})$ will be depressed in the presence of lattice
fluctuations that grow up below 
  $ T_{SP}^0 $ (Fig.~5.a ). In the very low-temperature domain, the SP 
correlation length becomes   exponentially large  and  the magnetic
susceptibility becomes thermally activated. When these temperature conditions
prevail, the system is almost long-range ordered.
 
{\epsfxsize 7cm
\begin{figure}
\centerline{\epsfbox{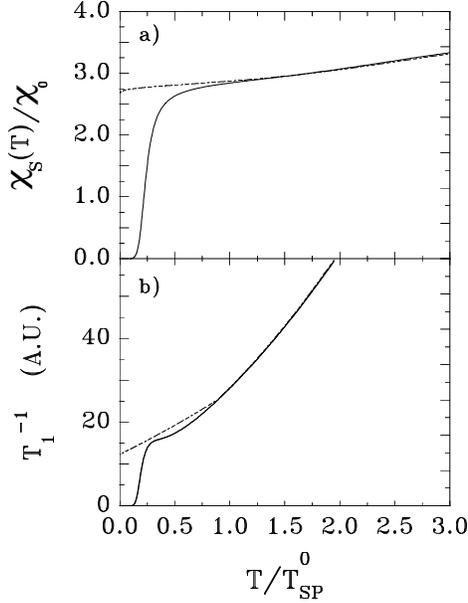}}
\caption{ Calculated  low temperature variation of a) the uniform
magnetic susceptibility and b) nuclear relaxation rate in the presence
(continuous line) and absence (dashed line)  of lattice fluctuations.}
\end{figure}}
\subsubsection{Nuclear relaxation rate}
From the above results, the  temperature
dependence of the nuclear spin-lattice relaxation rate $T_1^{-1}$ can be
readily calculated. The basic expression for  $T_1^{-1}$ in one dimension is
well known to be
\begin{eqnarray}
T_1^{-1}= \mid \bar{ A}\mid^2 T \int dQ { {\rm Im}\chi(Q,\omega) \over
\omega},
\end{eqnarray}
which can be taken in the limit $\omega \to 0$. Here  ${\rm Im}\chi$ is the
imaginary part of the retarded spin response function and $\bar{A}$ is a
constant  proportional to the hyperfine coupling \cite{moryia}. It is well
established that
 uniform $(Q\sim 0$) and  AF $(Q\sim 2k_F)$
spin fluctuations give the essential contributions to the
relaxation in one dimension\cite{CB2},  allowing in turn to make the
following decomposition
\begin{eqnarray}
T_1^{-1}= T_1^{-1}[Q\sim 0] + T_1^{-1}[Q\sim 2k_F].
\end{eqnarray}
For
the staggered part
$T_1^{-1}[Q\sim 2k_F]$, we have seen from (\ref{imchi}) that ${\rm
Im}\chi(2k_F + q,\omega)$ ( whose definition coincides with $ -{\rm
Im}\chi_{\mu \ne 0}^M(2k_F + q,\omega)$) is peaked when  $q$ lies in the
interval 
$[-T/v_F,T/v_F]$, which leads  to   
\begin{equation}
T_1^{-1}[Q\sim 2k_F] \simeq C_1 T 
D[T,\{\phi^-\}]\bar{\chi}^+_{\mu}(T,\{\phi^-\}),
\label{2kf}
\end{equation}
 where $C_1=\pi v_F^{-1}\mid \bar{ A}\mid^2\tanh(1/4) $.

 As for the uniform contribution, the use of (\ref{chis}-\ref{chi0}) immediately
leads to
\begin{equation}
T_1^{-1}[Q\sim 0] \simeq C_0 T { D[T,\{\phi^-\}]\over \bigl[1-{1\over 2}g_1(T)
D[T,\{\phi^-\}]\bigr]^2},
\label{q0}
\end{equation}
where $C_0= 4\pi (v_F)^{-1}\mid\bar{ A}\mid^2$.

As expected, the reduction of spin fluctuations by short range lattice
correlations below $T_{SP}^0$ will decrease the amplitude of the relaxation via
the reduction of the density of states.  At sufficiently low temperature,
namely when the correlation length $\xi_{SP}$ becomes exponentially large,
one finds $T_1^{-1} \sim \chi_s \sim e^{-\rho T_{SP}^0/ T}$. In the
 high-temperature regime, lattice fluctuations are small and the uniform
component
$T_1^{-1} \sim C_0T \chi^2_s$ eventually dominates the relaxation. Using the
scaling form (\ref{chibarfl2}) in  (\ref{2kf}) and (\ref{q0}), the
temperature dependence of
$T_1^{-1}$ is summarized in Figure 5.b where it is compared to the case without
fluctuations.

\section{Application to the SP  system (TMTTF)$_2$PF$_6$ }
  By way of application of the present theory to the  sulphur
based organic compound (TMTTF)$_2$PF$_6$. Previous analysis  of this system
in the normal phase using the one-dimensional electron gas model can be used for
the determination of the input  bare parameters of the purely electronic part of
the model. NMR analysis of Wzietek {\it et al.,}\cite{wzietek2} have shown that
$\gtil_1\simeq\gtil_2\simeq .9$, with
$E_F\simeq 1600$K, give a rather good description of  the temperature
dependent magnetic susceptibility in the high temperature domain of this
material. 
 As for the observed characteristic temperature scale $T_\rho\approx 220$K 
(below which the system presents insulating properties), it can be used to
identify, together with (\ref{couplings}), the low-temperature domain of strong
umklapp scattering, which then allows to take
$\gtil_3\approx.2$\cite{CaronCB}. As for the input parameters for the lattice
component, one will fix the value of $T_{SP}^0$ at $60$K which is the
characteristic temperature scale for the onset of strong lattice fluctuations in
x-ray  experiments\cite{pouget}. From the above set of 
figures  all quantities of interest can be calculated.

\subsection{Magnetic susceptibility}
The temperature-dependent EPR spin susceptibility (TMTTF)$_2$PF$_6$ measured by
Creuzet {\it et al.,}\cite{creuzet}    is reported in the low-temperature
domain in Figure~6 . 
$\chi_s(T)$ decreases monotonously from the high temperature domain and becomes
weakly temperature dependent near 80K, which is typical of   all members of the
sulphur series in the normal state. However, in the low temperature domain below 
$60$K,  the spin susceptibility  decreases by roughly 40 \%  
down to the true SP  transition at $T_{SP}\approx 19$K, below which
it becomes thermally activated. 
{\epsfxsize 7cm
\begin{figure}
\centerline{\epsfbox{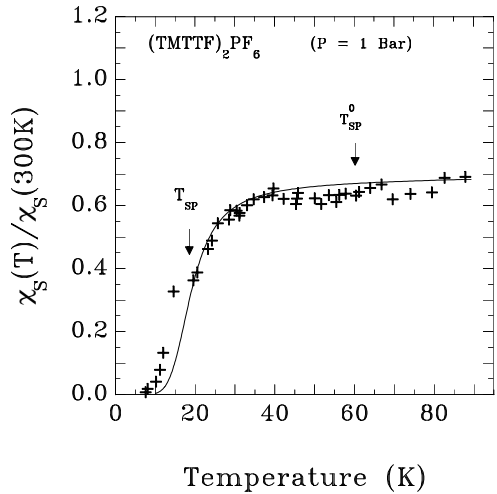}}
\caption{Comparison between calculated (continuous line) and observed
(crosses) temperature profiles of magnetic susceptibility for
(TMTTF)$_2$PF$_6$. The data are taken from Ref.[7].}
\end{figure}}
  Using the above set of parameters  for the  model, the theoretical prediction
for
$\chi_s(T,\{\phi^-\})$  is illustrated  in Figure~6  and gives a fairly good 
description of  the temperature variation of $\chi_s$ in the 
SP  pseudo-gap regime. At very low temperature, the  present theory 
predicts  a thermally activated behaviour when $\xi_{SP}$ grows
exponentially, which is found to     mimic the actual temperature dependence
below the true transition temperature at 19K.  However, a more realistic 
description of the SP  system in this low temperature  region, 
would require the inclusion of the interchain coupling.      

\subsection{Nuclear Relaxation}

 The temperature profile of $T_1^{-1}$ for (TMTTF)$_2$ PF$_6$,   measured  by
Creuzet et al.,\cite{creuzet} is given in Figure ~7. From the  analysis of
nuclear relaxation in the high temperature domain $T_{SP}^0 <\ T\ < T_\rho$,
where SP  fluctuations are weak,  the contribution to  
$T_1^{-1}$ is well known to be purely electronic in character. A quantitative
description of the relaxation rate in this regime  can be obtained
from (\ref{2kf}) and (\ref{q0}) neglecting the dependence on
$\phi^-$\cite{wzietek2}.   Below
$T_{SP}^0$,  the relaxation rate shows a  30\% decrease between $T_{SP}^0$ and
$ T_{SP}$ due to one-dimensional lattice fluctuations.  
According to (\ref{2kf}) and (\ref{2kf})  both the staggered and uniform
parts of the relaxation are affected below
$T_{SP}^0$. Thus, using these expressions for the above set of parameters,
one obtains the 
$T_1^{-1}$ temperature profile shown in  Figure~7. The theoretical curve is
obtained  from the expressions (\ref{q0}), (\ref{2kf}) and (\ref{chibarfl2})
in which  the values of the constants $C_0(\pi v_F)^{-1}\simeq 12.1$ and $C_1(\pi
v_F)^{-1}T_\rho\bar{\chi}^+_{\mu}(T_\rho)\simeq 9.6$ results from  the analysis
of
$T_1^{-1}$ data made by Wzietek {\it al.,} \cite{wzietek2} in the high
temperature domain
$T\gg T_{SP}^0$.  
{\epsfxsize 7cm
\begin{figure}
\centerline{\epsfbox{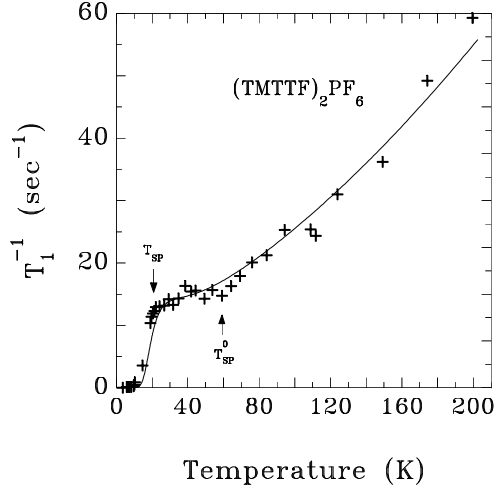}}
\caption{Comparison between calculated (continuous line) and observed
(crosses) temperature profiles of nuclear relaxation rate for
(TMTTF)$_2$PF$_6$. The data are taken from Ref.[7].}
\end{figure}}
\acknowledgements
The authors thank J.-P. Pouget and L. G. Caron for
 numerous discussions. We would also like to thank D. Senechal for
useful comments about the manuscript. Financial support from the Natural Sciences
and Engineering Research Council of Canada (NSERC), le Fonds pour la Formation
de Chercheurs et l'Aide
\`a la Recherche du Gouvernement du Qu\'ebec (FCAR)  and Canadian Institute
for Advanced Research (CIAR) is gratefully acknowledged.

\appendix
\section{Elementary Susceptibilities in presence of phonons} 
In this appendix, we  proceed to the  calculation of the
elementary  electron-hole bubbles    dressed by  lattice 
fluctuations in the Landau, Peierls and Cooper channels, respectively.
\subsection{Susceptibility at small $q$ and $\omega$}
 The expression corresponding to the electron-hole bubble of  Figure 3a
is
\begin{equation}
\chi(\qtil, \{\phi^-\})=  -{T\over L} \sum_{p,\ktil}
G_p(\ktil, \{\phi^-\}) G^0_p(\ktil +\qtil),
\end{equation}
for one spin orientation. Using the spectral representation of $G_p(\ktil,
\{\phi^-\})$ and performing the fermion frequency  sum, one finds
\begin{equation}
\chi(\qtil, \{\phi^-\})=  -(\pi L)^{-1}\sum_{p,k}\int_{-\infty}^{+\infty}
d\omega' {\rm Im }G_p(k,\omega', \{\phi^-\}){ n[\epsilon_p(k+q)] -
n[\omega']\over i\omega_m +\epsilon_p(k+q) -\omega'}.
\end{equation}
Since most of the spectral weight appears in the region
$\omega'
\approx
\epsilon_p(k)$, one can replace $\epsilon_p(k)$ by $\omega'$ in the ratio
 of the above integral, which can be cut off at $\pm E_0/2$. From the definition
of the density of states (\ref{dos}), one finds, after analytic
continuation to real frequencies,
\begin{eqnarray}
{\rm Re}\chi(q,\omega, \{\phi^-\})= && {1\over 2} \int_{-E_0/2}^{+E_0/2}
D[\omega',\{\phi^-\}] \left(-{\delta n\over \delta \omega'}\right) d\omega'\sum_p
{pv_Fq\over pv_Fq-\omega}\cr
{\rm Im}\chi(q,\omega, \{\phi^-\})= && {1\over 2}\int_{-E_0/2}^{+E_0/2}
D[\omega',\{\phi^-\}] \left(-{\delta n\over \delta \omega'}\right)
d\omega'\sum_p  p\pi v_Fq\delta(pv_Fq-\omega). 
\end{eqnarray}   
 Here $D(\omega,\{\phi^-\})$ is the density
of states per spin in the presence of the spin-Peierls pseudo-gap. 
The calculation of $\chi(\qtil, \{\phi^-\})$ at the step $\ell$ of the 
renormalization group procedure gives the same expression, except for  $E_0$
which is replaced by $ E_0(\ell)$.

\subsection{Peierls and Cooper susceptibilities}
The calculation of the Peierls electron-hole bubble of Figure 3b starts with the
following expression 
\begin{eqnarray}
\chi(2k_F +q, \omega_m, \{\phi^-\})= && {2T\over L} \sum_{k,\omega_n}
G_-(k,\omega_n,\{\phi^-\})G^0_+(k+2k_F +q,\omega_n+\omega_m),
\end{eqnarray}
for both spin orientations. At zero external variables 
and  after a summation over  the fermion frequencies and the use of the spectral
representation for $G_-(\ktil,\{\phi^-\})$, one gets
\begin{eqnarray} 
\chi(2k_F, \{\phi^-\})= -{2\over
\pi L}\sum_k\int_{-\infty}^{+\infty} d\omega'{\rm
Im}G_-(k,\omega,\{\phi^-\}){n[\omega']- n[-\epsilon_-(k)]\over
 \omega' + \epsilon_-(k)},
\end{eqnarray}
 which actually coincides with the real part of the Peierls bubble. Assuming that
${\rm Im} G_-(k,\omega',\{\phi^-\})$  is peaked in the region $\omega'\approx
\epsilon_-(k)$, from which one replaces  $\epsilon_-(k)$ by $\omega'$
in the ratio appearing in  the integral, we find
\begin{equation}
 \chi(2k_F,\{\phi^-\})= -\int_0^{E_0/2} D[\omega',\{\phi^-\}] d\omega'
{\tanh(\beta\omega'/2)\over \omega'}.
\end{equation}
As for the electron-electron (Cooper) elementary bubble at zero external
variables corresponding to  the expression 
\begin{eqnarray}
\chi(\{\phi^-\})= {2T\over L} \sum_{\ktil} G_-
(\ktil,\{\phi^-\})G^0_+(-\ktil),
\end{eqnarray}
the property $G^0_+(-k,-\omega_n)=-G^0_+(k+2k_F,\omega_n)$ leads to the
relation  $\chi(2k_F,\{\phi^-\})=-\chi(\{\phi^-\})$.  

Within the
renormalization group scheme at
$\ell$, the same expressions  become
\begin{eqnarray}
d\chi(2k_F,\{\phi^-\})= && - D[E_0(\ell)/2,\{\phi^-\}]d\ell \cr
            = && - d\chi(\{\phi^-\}), 
\end{eqnarray}
when evaluated in the outer energy shell. Finally, after analytic continuation to
real frequencies of  (A4), we can carry over the same type of  calculation for 
the  imagnary part of the Peierls bubble at $T$ in the limit of  small frequency
and we find 
\begin{eqnarray}
{\rm Im \chi}(2k_F +q, \omega\to 0,\{\phi^-\}) \simeq &&-{\pi\over 2}
\int_{-E_0/2}^{+E_0/2} d\omega' D[\omega',\{\phi^-\}]\bigl(n[\omega']-n[-\omega'
+ v_Fq]\bigr) \delta\bigl( \omega' + (\omega- v_Fq)/2\bigr)\cr
= && -{\pi D[v_Fq/2,\{\phi^-\}]
\over 8T\cosh^{2}(\beta v_{F}q/4)}\ \omega. 
\end{eqnarray}

\section{One-particle self-energy  and four-point vertex part in presence of
phonons}
\subsection{One-particle self-energy}
The expression for  the one particle self-energy  diagram  of Figure ~4a reads
\begin{eqnarray}
\Sigma_+(\ktil,\{\phi^-\})= && -2 g^2 {T^2\over L^2} \sum_{\ktil',\qtil}
G_-(\ktil',\{\phi^-\})G_-^0(\ktil' +\qtil)G^0_+(\ktil-\qtil) \cr
 \simeq && 2g^2 {T\over \pi L^2} \sum_{k'}\sum_{\qtil}\int_{-\infty}^{+\infty}
d\omega'{\rm Im
}G^(k',\omega',\{\phi^-\}){n[\omega']-n[\epsilon_-(k')-v_Fq]\over [i\omega_n-
\epsilon_+(k)- i\omega_m  +v_Fq][\omega'+ i\omega_m- \epsilon_-(k') + v_Fq]},
\end{eqnarray}
where the second line results from a fermion frequency summation and the use 
of the spectral weight representation. The approximation scheme of Appendix A
then allows to  put $\epsilon_-(k') \approx \omega'$ in the ratio appearing in 
r.h.s of this last expression and to cut off the integral over $\omega'$  at
$\pm E_0/2$.  In the RG procedure,   the outer
energy shell evaluation of $\Sigma_+$ at $\ell$, which is obtained after  the  
frequency sum over
$\omega_m$  leads to $d\Sigma_+ = d\Sigma_+^+  + d\Sigma_+^-$, where
\begin{eqnarray}
d\Sigma_+^\pm(\ktil,\{\phi^-\}) = && - {g^2\over 4\pi}
D[E_0(\ell)/2,\{\phi^-\}] \ dE_0(\ell) \cr
&& \ \ \ \ \ \ \ \ \ \times 
\int dq  {\bigl(n[\pm E_0(\ell)/2- v_Fq]-n[\pm
E_0(\ell)/2]\bigr)\bigl(n_B[-v_Fq] + n[v_Fq +
\epsilon_+(k)]\bigr)\over 2v_Fq +i\omega_n -\epsilon_+(k)}\cr
 \simeq && - {g^2\over 8\pi v_F} [G^0(\ktil)]^{-1}D[E_0(\ell)/2,\{\phi^-\}]\ 
d\ell,
\end{eqnarray}
to leading order in $[G^0(\ktil)]^{-1}= i\omega_n -\epsilon_+(k)$, for $\beta
E_0(\ell) \gg 1$, and where
$n_B[x]=(e^{\beta x}-1)^{-1}$. Here the integration over the momentum transfer
$q$ is found to contribute only  in the interval
$2v_Fq_0 > 2v_F\mid q\mid\  > E_0(\ell)$, where $q_0$ is a momentum
transfer cut-off. 

\subsection{Vertex part}

 The evaluation of the two-loop vertex part in the presence of lattice
fluctuations proceeds along similar lines. The diagrams of Figure~4b corresponds
to
$\Gamma^{(2)}_a + \Gamma^{(2)}_b $ and to the generic expression
\begin{eqnarray}
\Gamma^{(2)}_{a,b}(\{\phi^-\})= && -g^2 {T^2\over L^2}\sum_{\ktil',\qtil'}
G_\mp(\ktil',{\phi^-})G^0_\mp(\ktil'-\qtil')G_\pm^0(\ktil_{1,2}+ \qtil')
G^0_\pm(\ktil_{1,2}+\qtil'-\qtil).
\end{eqnarray} 
 In the RG sense one can drop   the
dependence on the external variables 
$\{\ktil_1,\ktil_2,\qtil\}$. Performing frequency sums on
$\omega_{n'}$ and $\omega_{m'}$ and using the above approximation scheme, one
gets 
\begin{eqnarray}
d\Gamma^{(2)}_{a,b}\simeq && g^2 {1\over 32\pi v_F^2 
 }D[E_0(\ell)/2,\{\phi^-\}]\ 
dE_0(\ell)
\cr
&& \times \int {dq'\over q'2}  \Bigl(\bigl[n[E_0(\ell)/2 + v_Fq']
-n[E_0(\ell)/2\bigr]\bigl[n_B[-v_Fq'] + n[v_Fq']\bigr] + E_0(\ell)\to
-E_0(\ell)\Bigr)\cr
\simeq && g^2{1\over 8\pi v_F }D[E_0(\ell)/2,\{\phi^-\} ]\ d\ell, 
\end{eqnarray}
for $2v_Fq_0 > 2v_F\mid q'\mid\  > E_0(\ell)$, $\beta v_Fq' \gg 1$, and $ \beta
E_0(\ell) \gg 1$.

\end{document}